\documentstyle[12pt,lathuile,epsfig]{article}
\begin{document}
\title{ 
MONOLITH: A HIGH RESOLUTION \\
NEUTRINO OSCILLATION EXPERIMENT
}
\author{
Tommaso Tabarelli de Fatis \\
{\em I.N.F.N. - Sezione di Milano} \\
{\em Piazza della Scienza 3, I-20126 Milano, Italy} \\
for the MONOLITH Collaboration                  \\
}
\maketitle
\baselineskip=14.5pt
\begin{abstract}
MONOLITH is a proposed massive magnetized tracking calorimeter
at the Gran Sasso laboratory in Italy, optimized for the detection
of atmospheric muon neutrinos.
The main goal is to test the neutrino oscillation hypothesis through 
an explicit observation of the full first oscillation swing. 
The $\Delta m^2$ sensitivity range for this measurement comfortably
covers the entire Super-Kamiokande allowed region.
Other measurements include studies of matter effects, the NC/CC and
$\bar{\nu}/\nu$ ratio with atmospheric neutrinos and auxiliary 
measurements from the CERN to Gran Sasso neutrino beam. 
Depending on approval, data taking with part of the detector could
start in 2005.  The MONOLITH detector and its performance are
described. 

\end{abstract}

\vspace{2cm}

\begin{center}
{\em Talk given at \\
     ``Les Rencontres de Physique de la Vall\'ee d'Aoste'', \\
     La Thuile (Italy), March 4-10$^{th}$, 2001}
\end{center}

\baselineskip=17pt
\newpage
\section{Introduction}

The current observations of atmospheric neutrinos\cite{SK,SK-2000}
are all consistent with a two-state neutrino oscillation in the
$\nu_\mu \to \nu_\tau$ channel. However, the experimental resolution
is too poor to clearly resolve the oscillation pattern and models with
non-standard $\nu_\mu$ disappearance
dynamics\cite{decay,decoherence,barbieri} are not ruled out by data. 
Moreover, scenarios involving admixtures of the electron neutrino
(3$\nu$) and of an hypothetical sterile  neutrino (4$\nu$), that would 
accommodate the current phenomenology, are not fully
constrained\cite{Fogli01} and even in the simple two-neutrino
oscillation picture, a wide range of oscillation parameters is still
allowed. 

MONOLITH\cite{Monolith} is a proposal for a detector explicitly 
designed to establish the occurrence of oscillations in atmospheric 
neutrinos and to discriminate alternative explanations, through the
observation of the full first oscillation swing in $\nu_\mu$
disappearance. This also yields a significantly improved measurement 
of the oscillation parameters.  
The occurrence of matter effects in atmospheric neutrino oscillations
can also be tested with MONOLITH, exploiting its capability to 
identify the lepton charge of muon neutrinos. 
This will constrain 3$\nu$ and 4$\nu$ scenarios and offer the unique
opportunity to test the neutrino mass hierarchy, before the advent of
$\nu$-factories. 

If approved, the detector will be located at the Gran Sasso Laboratory
in Italy, where the measurement of atmospheric neutrinos can be
supplemented by measurements in the CNGS beam from CERN.

\section{The MONOLITH detector}

\begin{figure}
\begin{center} 
\epsfig{file=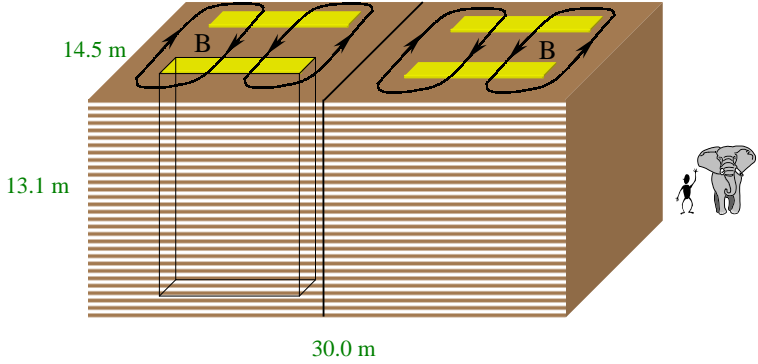, angle=-90, width=10cm}
\caption{\small Schematic view of the MONOLITH detector.
The arrangement of the magnetic field is  also shown.} 
\label{fig:module}
\end{center}
\end{figure}

MONOLITH is a massive tracking calorimeter with a coarse structure
and magnetic field. A large modular structure has been chosen for the
detector (figure \ref{fig:module}). One module  consists in a stack of
120 horizontal 8 cm thick iron planes with a surface area of
$14.5\times 15\ {\rm m^2}$, interleaved with 2 cm planes of  sensitive
elements. The height of the detector is about 13 meters. 
The magnetic field configuration is also shown in figure
\ref{fig:module}; iron plates are magnetized at a magnetic induction
of $\approx 1.3$ T. The detector consists  of two modules and its 
total mass exceeds 34 kt. 
  
Glass Spark Counters (resistive plate chambers with glass electrodes)
have been chosen as active detector elements. They provide two
coordinates with a pitch of 3 cm, and a time resolution of order 1~ns.
An external veto made of scintillation counters is foreseen to reduce
the background from cosmic ray muons. 

\section{Measurements with atmospheric neutrinos} 
\subsection{Test of $\nu_\mu$ disappearance dynamics} 

Oscillation studies with atmospheric neutrinos requires that the
energy $E$ and the baseline $L$ of the incoming neutrino be measured
in each event. The latter is related to the neutrino direction and 
may be estimated in charged-current (CC) interactions from the direction 
of the outgoing lepton. MONOLITH, with a mass comparable to
Super-Kamiokande, has a much larger acceptance to muons at high
energies, where the muon direction gives a better estimate of the
neutrino direction. This results in a considerably improved $L/E$
resolution and overcomes the main limitation in the sensitivity to the
oscillation pattern of current atmospherics neutrino experiments.

High energy atmospheric neutrinos bear another advantage: they 
represent an ideal source for a disappearance experiment, since 
they naturally comprise the {\it near} (down-going neutrinos) and a
{\it far} (up-going neutrinos) positions\footnote{MONOLITH has an 
effective threshold on $\nu_\mu$-CC events around 3~GeV, where the
asymmetry of atmospheric neutrino fluxes, mainly due to geomagnetic
effects, is below the percent level \cite{Gai98}.}. 
Thus a detailed prior knowledge of neutrino fluxes is not required.
The atmospheric neutrino beam is also characterized by a wide range 
in $L/E$ (from about 1 km/GeV to 10$^5$ km/GeV), that  gives access 
to a very large range of oscillation parameters and, in particular,
results in a unique sensitivity to oscillations if $\Delta m^2$ is
low.

\begin{figure}[t]
\begin{center}
\begin{tabular}{c}
\mbox{\epsfig{file=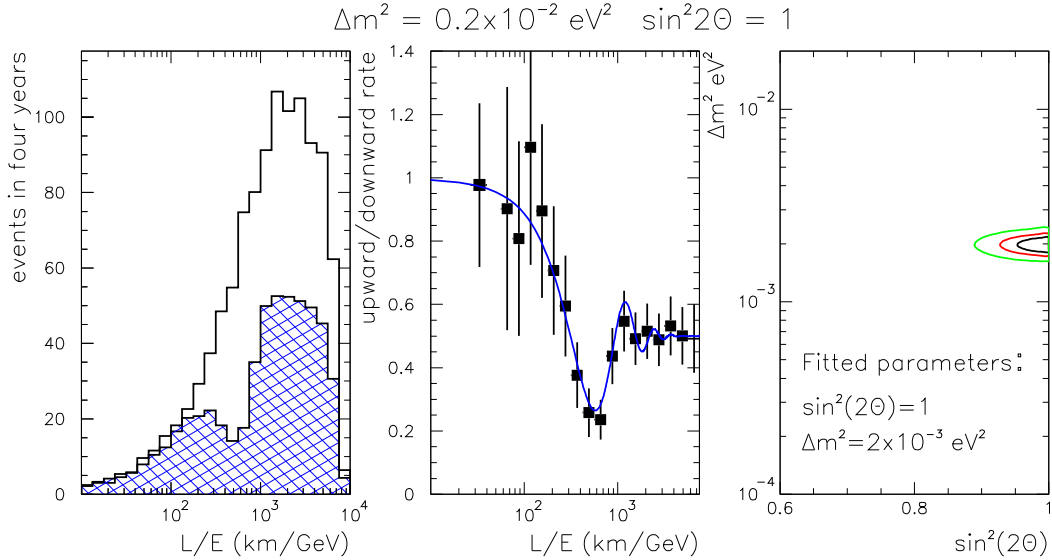,width=0.95\textwidth}} \\
\end{tabular}
\end{center}
\caption{\small Results of the $L/E$ analysis on a simulated sample
  in the presence of $\nu_\mu\rightarrow\nu_\tau$   oscillations, with
  parameters $\Delta m^2 = 2\times 10^{-3}$ eV$^2$ and $\sin^2
  (2\Theta)=1.0$. 
  The figure shows from left to right: %
  the $L/E$ spectrum of up-going muon neutrino events (hatched area) and
  the reference spectrum of down-going muon neutrino events, which 
  are assigned the baseline $L$ they would have travelled if 
  they were produced with a nadir angle equal to the observed zenith
  angle \cite{Pic97} (open area); 
  their ratio with the best-fit of $\nu_\mu \to \nu_\tau$ oscillation 
  superimposed; the result of the fit with the corresponding allowed 
  regions for oscillation parameters at 68\%, 90\% and 99\%
  C.L.. The simulated statistics correspond to 25 years of exposure, 
  errors and sensitivity contours correspond to four years.}
\label{fig:one}
\end{figure}

Figure \ref{fig:one} shows the expected $L/E$ distributions of
up-going neutrinos and of the reference sample of down-going
neutrinos and their ratio, for $\nu_\mu \to \nu_\tau$ oscillation 
with parameters around the current best-fit to Super-Kamiokande data. 
The figure also shows the allowed regions of the oscillation parameter
space of the experiment after four years of exposure. 
 
\begin{figure}[t]
\begin{center}
\mbox{\epsfig{file=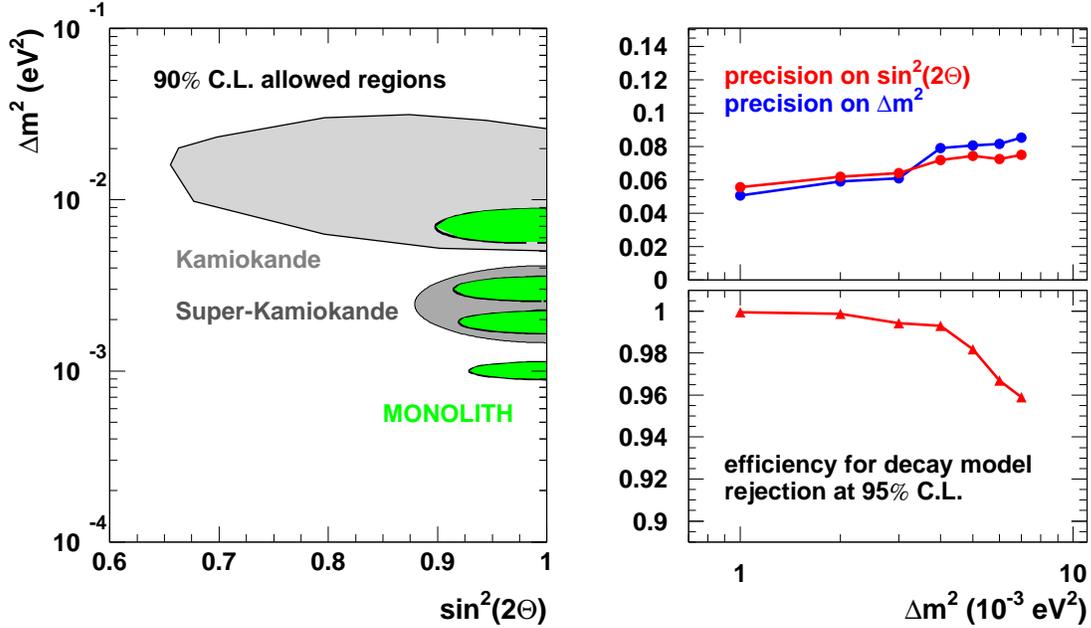,width=.98\textwidth}}
\end{center}
\caption{\small {\it Left:} 
  Expected allowed regions of $\nu_\mu-\nu_\tau$ oscillation
  parameters for MONOLITH after four years of atmospheric neutrino
  exposure. The results of the simulation  for $\Delta
  m^2 = 1, 2, 3, 7 \times 10^{-3}$ eV$^2$ and maximal mixing are
  shown. The 90\% allowed regions for $\nu_\mu \to
  \nu_\tau$ oscillations of Super-Kamiokande \cite{SK-2000} and
  Kamiokande \cite{Kamiokande}  experiments are also shown.
  {\it Right}: MONOLITH expected precision on oscillation parameters
  and discrimination power against the decay model of
  Ref. \cite{decay}. 
}
\label{sensibilita}
\end{figure}

An oscillation pattern is visible, at clear variance with the
unconventional interpretations \cite{decay,decoherence,barbieri}  
of Super-Kamiokande data, which share the feature of the absence of a
first clear dip in the $L/E$ distribution. A detailed analysis reveals 
that if the true process is oscillations, the alternative hypotheses 
can be rejected with more than 99\% efficiency at 95\% C.L., over the 
entire region of parameters allowed by Super-Kamiokande (see
Fig. \ref{sensibilita}). The clearness of the oscillation pattern also 
results in a significantly improved measurement of the oscillation 
parameters. A precision on $\Delta m^2$ better than 10\% is
anticipated (Fig. \ref{sensibilita}).

\subsection{Detection of matter induced effects}

The very long baselines available with atmospheric neutrinos offer the
possibility to search for Earth-induced matter effects. In that
endeavour, atmospheric neutrino experiments are not contested by
current and planned conventional accelerator beam experiment, whose
baselines are too short for a significant effect.

Matter effects can play an important role if there are significant
contributions of $\nu_e$ or $\nu_{\rm sterile}$ to atmospheric
neutrino oscillations. The non-observation of large matter effects 
has been already exploited by Super-Kamiokande to 
exclude the pure $\nu_\mu - \nu_{\rm sterile}$ oscillation hypothesis
at  99\% C.L. \cite{SK,SK-2000}. For a contribution of non-maximal
$\nu_\mu - \nu_{\rm sterile}$ oscillations, matter effects would also
manifest themselves in differences in the oscillation patterns for
neutrinos and anti-neutrinos. Such differences could be measured with
MONOLITH \cite{Monolith}, thanks to its capability to identify 
the muon charge, and could yield important constraints on hybrid
oscillation scenarios \cite{Lisi,yasudahybrid}. 

Such effects could be detectable even in standard three flavour
oscillation scenarios \cite{Banul}. A subdominant $\nu_e$ mixing could
sizeably modify the $\nu_\mu$ transition probability in particular
regions of phase space where the $\nu_\mu \to \nu_e$ transition
becomes resonant in matter. 
Depending on the sign of $\Delta m^2$, such effects occur
either for neutrinos or for anti-neutrinos only. By comparing the
neutrino and anti-neutrino distributions in MONOLITH, the sign of
$\Delta m^2$, and therefore the neutrino mass hierarchy, could be
determined if a signal would be observed. 

\begin{figure}[t]
\begin{center}
\mbox{\epsfig{file=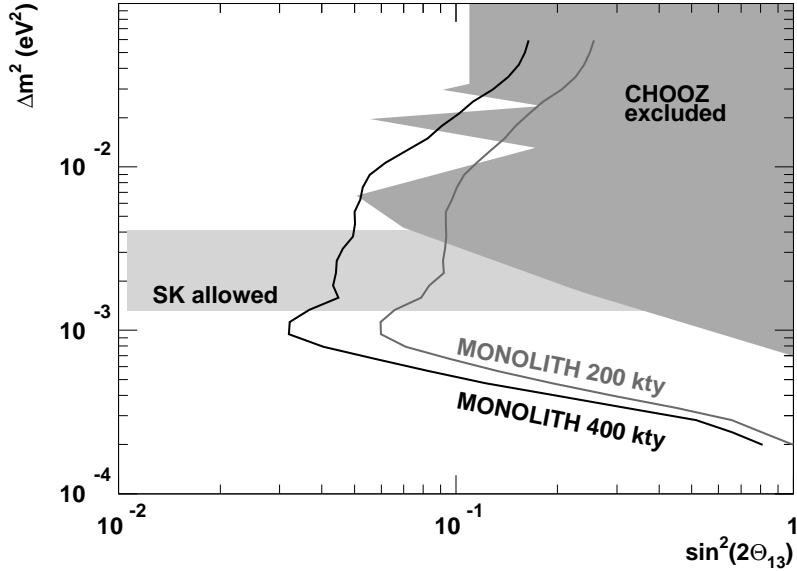,width=11cm}}
\end{center}
\caption{\small
  Region of oscillation parameters in the $3\nu$ scenario over which
  the sign of $\Delta m^2$ can be determined at 90\% C.L. after 
  200~kty and 400~kty of MONOLITH exposure. The regions excluded by
CHOOZ results and allowed by Super-Kamiokande data are also shown.}
\label{matm}
\end{figure}

Figure \ref{matm} shows a preliminary estimate of the region of
oscillation parameters over which the sign of $\Delta m^2$ can be
determined at 90\% C.L. after 200~kty of MONOLITH exposure. The
sensitivity achievable in a long-term run (or with a larger detector),
corresponding to an exposure of 400~kty, is also shown. For
comparison, the region excluded by CHOOZ \cite{CHOOZ} and the region 
allowed by Super-Kamiokande are displayed. This analysis assumes 
one mass scale dominance for atmospheric neutrinos (i.e. $\Delta
m^2=m^2_3 - m^2_{1,2}$) and maximum mixing in the (2,3) sector, as
suggested by Super-Kamiokande data, and no prior knowledge of
$\sin^2(2\Theta_{13})$. The latter parameter can also be constrained 
by MONOLITH over a region that will be partly accessible to MINOS
\cite{Para} and fully accessible to long term low-energy beam projects
searching for the subdominant $\nu_\mu \to \nu_e$ transition
\cite{JHF}. These experiments, however, can not measure the sign
of $\Delta m^2$.  

\section{Measurements with neutrino beams}
\subsection{Oscillation studies with the CNGS beam}

The CNGS beam \cite{CNGS} from CERN to Gran Sasso is optimized for
$\nu_\tau$ appearance experiments and the energy spectrum is somewhat 
too high to allow the detection of the first entire oscillation
swing in a disappearance experiment. Nonetheless, MONOLITH can
reconstruct with high efficiency and good resolution about 40000
$\nu_\mu$ charged-current interactions per year. The systematic error, 
mainly related to the knowledge of the beam spectra, need to be
suitably controlled for a disappearance measurement on the CNGS
beam. This is partly achieved in the measurement of the neutral to 
charged current ratio (NC/CC). In this method looser requirements on
the muon reconstruction may be applied and about 100000 events per
year (CC+NC) can be reconstructed and classified according their 
length and shape. Both these method can be used to complement the
disappearance measurements with atmospheric neutrinos, in particular
if $\Delta m^2$ is high.  

\subsection{Measurements at a neutrino factory}

In the long term, MONOLITH could be used as a target detector in a 
potential neutrino factory beam. Neutrino beams from future muon
storage rings \cite{Blondel} (neutrino factories) will be essentially
pure beams of either $\nu_\mu + \bar\nu_e$ or $\bar\nu_\mu + \nu_e$. 
The occurrence of $\nu_e - \nu_\mu$ or $\nu_e -\nu_\tau$ oscillations
would manifest itself via the appearance of wrong sign muons.
It has been checked that MONOLITH, with good muon charge separation 
and momentum measurement, is well suited for the observation of such
oscillations.  

Neutrino factories will in particular offer the possibility to measure
the $\Theta _{13}$ mixing angle, the sign of $\Delta m^2$ through
matter effects if not observed earlier and, depending on which of the
solar neutrino solutions is correct, it might also open the way for
the study of CP violation in the neutrino system. 

\section{Conclusions}

MONOLITH is a 34 kt magnetized iron tracking calorimeter proposed for
atmospheric neutrino measurements at the Gran Sasso Laboratory in Italy.
Its main goal is the proof of the neutrino oscillation hypothesis through
the explicit observation of a sinusoidal oscillation pattern.
Other goals include tests of the neutrino mass hierarchy in $3\nu$ and
$4\nu$ scenarios, trough the investigation of potential matter effects,
and auxiliary measurements in the CERN to Gran Sasso beam. 
In the long term, the detector could also be used in a potential
neutrino factory beam. 
\end{document}